# Effect of quantum group invariance on trapped Fermi gases

Marcelo R. Ubriaco*
*Laboratory of Theoretical Physics*
*Department of Physics*
*University of Puerto Rico*
*P. O. Box 23343, Río Piedras*
*PR 00931-3343, USA*

**Abstract**

We study the properties of a thermodynamic system having the symmetry of a quantum group and interacting with a harmonic potential. We calculate the dependence of the chemical potential, heat capacity and spatial distribution of the gas on the quantum group parameter $q$ and the number of spatial dimensions $D$. In addition, we consider a fourth-order interaction in the quantum group fields $\Psi$, and calculate the ground state energy up to first order.

*E-mail:ubriaco@ltp.upr.clu.edu

## 1 Introduction

The recent observations of Bose-Einstein condensation [1, 2, 3] has triggered numerous theoretical investigations on the behavior of a trapped Bose gas under experimental conditions. Furthermore, the stable fermionic isotope of lithium, $^6$Li, has been trapped and cooled in a similar way as its bosonic counterpart $^7$Li [4]. One of the most interesting aspects of a trapped Fermi gas is that, as reported in References [5, 6], for a sufficiently strong trapping magnetic field a BCS transition is expected to occur at densities and temperatures experimentally attainable. The critical temperature and one-particle excitations for the superfluid phase transition have also been studied [7, 8]. Two recent articles [9, 10] on trapped Fermi gases, in which the BCS transition is excluded, include a study on their thermodynamic behavior in harmonic traps.

The purpose of this paper is to study the behavior of a thermodynamic system having the symmetry of a quantum group. The role of quantum groups in physics has its origin in the quantum inverse scattering method [11], and the possible relevance they may play in other fields of physics has been the focus of great interest ever since. A quantum group, as compared with a classical group, contains an additional parameter $q$. In the quantum inverse scattering method the parameter $q$ acquires a physical meaning through its relation with the Planck's constant. In two recent articles [12, 13] we considered the system represented by the hamiltonian $\mathcal{H}_B = \sum_{i,\kappa} \overline{\Phi}_i(\kappa)\Phi_i(\kappa)$, $i = 1, 2$, with the operators $\Phi_i$ and the adjoint $\overline{\Phi}_i$ satisfying algebraic relations which are covariant under the quantum group $SU_q(2)$. For $q = 1$, the operators $\overline{\Phi}_i$ and $\Phi_i$ become standard creation and annihilation boson operators respectively. In this model, Bose-Einstein condensation is realized as a second order phase transition, and the heat capacity becomes discontinuous at the critical temperature in one, two and three dimensions, even without the presence of an external potential. In a similar fashion, one can define [14] a set of operators $\Psi_i$ and $\overline{\Psi}_i$, with corresponding quantum group covariant algebraic relations, such that for $q = 1$ they become standard fermionic operators.

A quantum group invariant hamiltonian is, at the moment, a mathematical model with phenomenological implications that depart from those predicted by either boson or fermion gases. This departure is measured by the value of the parameter $q$. There is, at present, no phenomenological evidence that quantum group symmetries are realized in any particular thermodynamic system. Since the operators $\Psi$ satisfy the Pauli exclusion principle and their representation in terms of ordinary fermions $\psi$ is not linear, any hamiltonian written in terms of them can be interpreted as an interacting fermion system. Our main motivation focuses in investigating about the role that these interactions, required by quantum group invariance, may play at low temperatures. The interaction term is fixed by the quantum group and its strength is parameterized by $q$. Naturally, for $q = 1$, these interactions will vanish and our results will reduce to the case of a gas with two fermion species interacting with a harmonic potential. Within the context of our results, only experiments with trapped Fermi gases can tell us whether quantum group symmetry is realized in the behavior of real gases. A comparison with the results reported in this paper will indicate whether a particular, and presumably small, deviation of the parameter $q$ from the standard value $q = 1$ fits the experimental data. Therefore, the calculations considered here, will help to answer questions regarding the possible relevance of quantum groups in a thermodynamic system. In this paper we study the thermodynamics of a system composed of quantum group fermions $\Psi_i$ interacting with a harmonic potential. For matter of simplicity, we will consider the case of $SU_q(2)$.

This paper is organized as follows. Section 2 is a very brief introduction to the formalism we will use in the following sections. A detailed discussion on quantum groups can be found in References [15]. In Section 3 we calculate the properties of this model interacting with an isotropic harmonic oscillator. In particular, we calculate the chemical potential, internal energy, heat capacity, and the particle distribution at $T = 0$ for an arbitrary number of spatial dimensions $D$. In Section 4 we introduce a fourth order $\Psi$-interaction and calculate the ground state energy up to first order. In Section 5 we summarize our results and compare them with the $q = 1$ case.

# 2 Free $SU_q(2)$ fermion gas

The quantum group $GL_q(2, C)$ consists of the set of matrices $T = \begin{pmatrix} a & b \\ c & d \end{pmatrix}$ with noncommuting elements generating the algebra $A_q$ [16]

$$\begin{aligned} ab = q^{-1}ba \quad &, \quad ac = q^{-1}ca \\ bc = cb \quad &, \quad dc = qcd \\ db = qbd \quad &, \quad da - ad = (q - q^{-1})bc, \end{aligned} \tag{1}$$

with the quantum determinant

$$det_q T = ad - q^{-1}bc. \tag{2}$$

The quantum determinant is defined such that it belongs to the center of the algebra. The relations in Equation (1) have the remarkable property that if the elements $a, b, c, d$ of $T$ commute with the elements $a', b', c', d'$ of $T'$, then the elements of the matrix $TT'$ also generate $A_q$. By setting $det_q T = 1$ and the unitary conditions [17]

$$\bar{a} = d \ \ , \ \ \bar{b} = -q^{-1} c, \tag{3}$$

leads to the quantum group $SU_q(2)$ with $q \in R$. It is clear that the quantum group $SU_q(2) \to SU(2)$ as the parameter $q \to 1$. In Reference [14] we introduced a set of operators $\Psi_i$ that transform under $SU_q(2)$ and become ordinary fermions $\psi_i$ in the $q \to 1$ limit. This is accomplished by the set of relations

$$\begin{aligned}
\{\Psi_2, \overline{\Psi}_2\} &= 1, \\
\{\Psi_1, \overline{\Psi}_1\} &= 1 - (1 - q^{-2}) \overline{\Psi}_2 \Psi_2, \\
\Psi_1 \Psi_2 &= -q \Psi_2 \Psi_1, \\
\overline{\Psi}_1 \Psi_2 &= -q \Psi_2 \overline{\Psi}_1, \\
\{\Psi_1, \Psi_1\} = \ & 0 \ = \{\Psi_2, \Psi_2\}.
\end{aligned} \tag{4}$$

Equations (4) are covariant under the action of the group $SU_q(2)$

$$\Psi_i = \sum_{j=1}^{2} T_{ij} \Psi_j, \tag{5}$$

and they become the $SU(2)$ covariant fermion algebra for $q = 1$. According to Equations (4), the operators $\Psi_j$ have a representation in terms of ordinary fermions $\psi_i$ as follows

$$\Psi_1 = \psi_1 \left[ 1 + (q^{-1} - 1) \psi_2^\dagger \psi_2 \right] \ \ , \ \ \Psi_2 = \psi_2, \tag{6}$$

with the corresponding equations for the adjoint $\bar{\Psi}_i$. Therefore, the simplest quantum group invariant hamiltonian $\mathcal{H} = \sum_\kappa \overline{\Psi}_\kappa \Psi_\kappa$ is written in terms of fermion operators as follows

$$\mathcal{H} = \sum_{\kappa} E_{\kappa} \left[ \psi^{\dagger}_{1,\kappa}\psi_{1,\kappa} + \psi^{\dagger}_{2,\kappa}\psi_{2,\kappa} + (q^{-2}-1)\psi^{\dagger}_{1,\kappa}\psi_{1,\kappa}\psi^{\dagger}_{2,\kappa}\psi_{2,\kappa} \right] \tag{7}$$

The grand partition function is given by

$$\mathcal{Z} = \prod_{\kappa} (1 + 2e^{-\beta(E_{\kappa}-\mu)} + e^{-\beta(E_{\kappa}(q^{-2}+1)-2\mu)}), \tag{8}$$

which for $q = 1$ becomes the partition function for a gas with two fermion species. The occupation number $\langle n \rangle$ is a function of the energy according to

$$\langle n(E) \rangle = \frac{1 + e^{-\beta E q^{-2}} z}{f(z,q)}, \tag{9}$$

where $z$ is the fugacity and the function

$$f(z,q) = e^{\beta E} z^{-1} + 2 + e^{-\beta E q^{-2}} z.$$

For $q > 1$ the function $\langle n(E) \rangle$ is very similar to the Fermi function. The chemical potential at zero temperature is independent of the number of space dimensions $D$

$$\mu_0(q > 1) = \frac{q^{-2}+1}{2} \mu_0, \tag{10}$$

where $\mu_0$ is the Fermi energy for $q = 1$. For $q < 1$, $\langle n(E) \rangle$ departs considerably from the Fermi case. At $T = 0$, states with energies $0 \leq E \leq q^2 \mu_0(q)$ are fully occupied, and those with energies $q^2\mu_0(q) \leq E \leq \mu_0(q)$ have occupation numbers $\langle n \rangle = 1/2$. The chemical potential, at $T = 0$ and $q < 1$, is related to the Fermi energy for $q = 1$ as follows

$$\mu_0(q < 1) = \left( \frac{2}{1+q^D} \right)^{2/D} \mu_0. \tag{11}$$

The total number of particles, in the thermodynamic limit, is given by the integral

$$\langle N \rangle = 2 \int_0^{\infty} g(E) \langle n(E) \rangle dE, \tag{12}$$

where $g(E)$ is the density of states.

## 3 $SU_q(2)$ fermion gas in a harmonic oscillator

The single-particle levels, excluding the zero-point energy, are given by the well known formula

$$E = \hbar\omega \sum_{j=1}^{D} n_j. \tag{13}$$

As an approximation, , we consider a continuous spectrum with density of states

$$g(E) = \frac{E^{D-1}}{(\hbar\omega)^D (D-1)!}. \tag{14}$$

A more accurate expression for the density of states was provided in References. [18] and applied to the Bose-Einstein condensation of a finite number of bosons in a harmonic oscillator trap, and in Reference [10] for the study of a free fermion gas interacting with a harmonic oscillator potential. Equation (14) will be sufficient for our purposes.

### 3.1 Chemical potential

The chemical potential, for low temperatures, can be obtained by splitting the integral in Equation (12) in the following intervals

a) $q < 1$: $[0, q^2\mu(q)]$, $[q^2\mu(q), \mu(q)]$, $[\mu(q), \infty]$,

b) $q > 1$: $[0, 2\mu(q)/(1+q^{-2})]$, $[2\mu(q)/(1+q^{-2}, q^2\mu(q)]$ , $[q^2\mu(q), \infty]$,

such that reverting the solutions leads to the results

$$\begin{aligned} \frac{\mu(q<1)}{\mu_0} &= \left(\frac{2}{1+q^{2D}}\right)^{1/D} - \ln 3 \, \frac{1-q^{2D}}{1+q^{2D}} \frac{kT}{\mu_0} \\ &+ (D-1)\left(\frac{1+q^{2D}}{2}\right)^{1/D} \left(\ln^2 3 \left(\frac{1-q^{2D}}{1+q^{2D}}\right)^2 - 2.88\right) \frac{1}{2} \left(\frac{kT}{\mu_0}\right)^2, \end{aligned}$$

$$\frac{\mu(q>1)}{\mu_0} = \frac{1+q^{-2}}{2}\left[1-(D-1)\frac{1.64}{(1+q^{-2})^2}\left(\frac{kT}{\mu_0}\right)^2\right]. \tag{15}$$

A simple inspection shows that for $q < 1$ the chemical potential depends also linearly on the temperature. In contrast to the fermion case, where $\mu$ is constant for $D = 1$, the function $\mu(q < 1)$ decreases with $T$. Figure 1 shows a graph of the chemical potential for $q = 0.2, 1, 2$, obtained from a numerical calculation of Equation (12). The chemical potential at zero temperature, $\mu_0(q)$, is independent of $D$ for $q > 1$. For $q < 1$, $\mu_0(q)$ decreases as $D$ increases. For low temperatures, the function $\mu(q < 1)$ depends more strongly on $q$ as we reduce the number of dimensions $D$.

## 3.2 Internal energy and heat capacity

The internal energy is obtained from the partition function

$$\begin{aligned} U &= -\frac{\partial \ln \mathcal{Z}}{\partial \beta} + \mu\langle N\rangle \\ &= \int_0^\infty E\; g(E)\; \frac{2+z(1+q^{-2})e^{-\beta q^{-2}E}}{f(z,q)}\; dE. \end{aligned} \tag{16}$$

The integral in Equation (16) can be solved for low temperatures by using the same kind of approximation we used to obtain the chemical potential. For $q \neq 1$ the internal energy is given by the equations

$$\begin{aligned} \frac{U(q<1)}{\langle N\rangle} &= \frac{D}{2}\mu_0\frac{2^{(D+1)/D}}{(D+1)(1+q^{2D})^{1/D}} \\ &+ \frac{D}{2}\mu_0\left(2.88-\frac{(1-q^{2D})^2}{(1+q^{2D})^2}\ln^2 3\right)\frac{(1+q^{2D})^{1/D}}{2^{1/D}}\left(\frac{kT}{\mu_0}\right)^2 \end{aligned} \tag{17}$$

$$\frac{U(q>1)}{\langle N\rangle} = \frac{D}{2}\mu_0\left[\frac{1+q^{-2}}{D+1}+\frac{1.64}{1+q^{-2}}\left(\frac{kT}{\mu_0}\right)\right]. \tag{18}$$

Figure (2) shows a graph of the heat capacity

$$\begin{aligned} C &= \frac{\partial U}{\partial T} \\ &= (D+1)\frac{U}{T} + z\beta\frac{\partial U}{\partial z}\left(\frac{\partial \mu}{\partial T} - \frac{\mu}{T}\right) \end{aligned} \tag{19}$$

for $D = 3$ and $q = 0.2, 1, 2$ which results from a numerical calculation of Equation (19).

### 3.3 Particle distribution at $T = 0$

In the semiclassical approximation, Thomas-Fermi approximation, the spatial distribution is given by the integral

$$n(r,T) = \frac{2\pi^{D/2}}{\Gamma(D/2)}\left(\frac{1}{2\pi\hbar}\right)^{D} 2\int_0^\infty \langle n(p,r)\rangle p^{D-1}dp. \tag{20}$$

Replacing Equation (9) and defining the variable $K = p^2/2m$, this integral becomes

$$n(r,T) = \frac{2\pi^{D/2}}{\Gamma(D/2)}\frac{(2m)^{D/2}}{(2\pi\hbar)^D}\int_0^\infty K^{(D-2)/2}\frac{e^{\beta(\hat{\mu}-K)} + e^{\beta(\hat{\mu}+\tilde{\mu}-(1+q^{-2})K)}}{1 + 2e^{-\beta(K-\hat{\mu})} + e^{\beta(\hat{\mu}+\tilde{\mu}-(1+q^{-2})K)}}dK, \tag{21}$$

where for convenience we have defined

$$\hat{\mu} = \mu(q) - \frac{1}{2}m\omega^2 r^2, \tag{22}$$

and

$$\tilde{\mu} = \mu(q) - \frac{q^{-2}}{2}m\omega^2 r^2. \tag{23}$$

For $q > 1$ we have $\tilde{\mu}_0 > \hat{\mu}_0$, and Equation (21) for $T = 0$ can be easily calculated if we divide the integral in the following intervals:

$[0, \hat{\mu}_0], [\hat{\mu}_0, (\hat{\mu}_0 + \tilde{\mu}_0)(1 + q^{-2})], [(\hat{\mu}_0 + \tilde{\mu}_0)(1 + q^{-2}), q^2\tilde{\mu}_0]$ and $[q^2\tilde{\mu}_0, \infty]$,

leading to the result

$$n(r,0) = 2\Lambda \left(1 - \frac{r^2}{r_F^2}\right)^{D/2}, \; q > 1, \tag{24}$$

where

$$\Lambda = \frac{2\pi^{D/2}}{\Gamma(D/2)} \frac{(D-1)!\langle N\rangle}{2\pi^D r_F^D},$$

and $r_F = \sqrt{2\mu_0/m\omega^2}$ is the Fermi radius for $q = 1$. This equation shows that the spatial distribution is independent of the parameter $q$ for all the values $q \geq 1$. A numerical calculation of Equation (21) shows that this independence on $q$ remains valid for $T > 0$.

For $q < 1$, a similar calculation leads to the result

$$n(r,0) = \Lambda \sqrt{\frac{2}{1+q^{2D}}} \left[ \left( q^2 - \frac{r^2}{r_F^2} \left(\frac{1+q^{2D}}{2}\right)^{1/D} \right)^{D/2} + \left( 1 - \frac{r^2}{r_F^2} \left(\frac{1+q^{2D}}{2}\right)^{1/D} \right)^{D/2} \right], \tag{25}$$

for $r < \left((2/(1+q^{2D})\right)^{1/2D} q r_F$.

$$n(r,0) = \Lambda \sqrt{\frac{2}{1+q^{2D}}} \left( 1 - \frac{r^2}{r_F^2} \left(\frac{1+q^{2D}}{2}\right)^{1/D} \right)^{D/2}, \tag{26}$$

for $r \geq \left((2/(1+q^{2D})\right)^{1/2D} q r_F$. From this results we see that the gas is less confined for $q < 1$. This fact becomes an expected result after rewriting the hamiltonian in terms of fermion operators. The fermion interaction term induced by the requirement of quantum group invariance becomes more repulsive as we decrease the value of $q$ below $q = 1$.

## 4 Higher order interactions

In the previous section we studied the simplest quantum group invariant system interacting with a harmonic potential. According to equations (6), introducing interactions between the operators $\Psi_i$ will lead to higher order interactions terms when the hamiltonian is rewritten in terms of ordinary fermion operators. We consider the simple interaction hamiltonian

$$\mathcal{H}_I = g \sum_{p'_1, p_1, p'_2, p_2} \overline{\Psi}_{p'_1,1} \overline{\Psi}_{p'_2,2} \Psi_{p_2,2} \Psi_{p_1,1}, \tag{27}$$

where $\{\Psi_{p',i}, \Psi_{p,j}\} = 0$, $\forall i, j$ and $p' \neq p$, and quantum group matrix elements $(a(p_i), b(p_i), c(p_i), d(p_i))$ commute with the set $(a(p_j), b(p_j), c(p_j), d(p_j))$ for $i \neq j$. In equation (27) the coupling $g$ is related to the scattering length $a$ by

$$g = \frac{4\pi a \hbar^2}{V m}, \tag{28}$$

and the summation is over all momenta satisfying

$$\mathbf{p'_1} + \mathbf{p'_2} = \mathbf{p_1} + \mathbf{p_2}. \tag{29}$$

The first-order correction to the energy of the system is given by those terms with $\mathbf{p'_1} = \mathbf{p_1}$ and $\mathbf{p'_2} = \mathbf{p_2}$. Again, it is convenient to rewrite the hamiltonian in terms of ordinary fermions. Replacing Equation (6) in Equation (27) leads to the first-order correction for the energy

$$E^{(1)} = g \frac{\langle N \rangle^2}{4} + g(q^{-2} - 1) \frac{\langle N \rangle}{2} \sum_p N_{p,1} N_{p,2}, \tag{30}$$

where $N_{p,i} = \psi_i^\dagger \psi_i$ and we took the equilibrium values

$$\sum_p N_{p,i} = \frac{\langle N \rangle}{2}. \tag{31}$$

In the ground state, all the fermions occupy the states with lower momenta, such that the sum $\sum_p N_{p,1}N_{p,2} = \langle N\rangle/2$. Thus, for the ground state first order correction we obtain the simple expression

$$E_0^{(1)} = gq^{-2}\frac{\langle N\rangle^2}{4}. \quad (32)$$

Equation (32) shows that the first order correction to the ground state energy decreases as we increase the value of the parameter $q$. From Equations (17) and (18) for $T = 0$ we write, up to first order, for the ground state energy $E_0 = E_0^{(0)} + E_0^{(1)}$

$$E_0(q < 1) = \frac{D(D!)^{1/D}}{D+1}\frac{\langle N\rangle^{1+1/D}}{(1+q^{2D})^{1/D}}\hbar\omega + gq^{-2}\frac{\langle N\rangle^2}{4} \quad (33)$$

$$E_0(q > 1) = \frac{D(D!)^{1/D}}{D+1}\frac{\langle N\rangle^{1+1/D}}{2^{1+1/D}}(1+q^{-2})\hbar\omega + gq^{-2}\frac{\langle N\rangle^2}{4} \quad (34)$$

Figure 4 shows the ground state function $\varepsilon_0 = 4E_0/g\langle N\rangle^2$ for $D = 3$. We have assumed the case of a dilute gas, $na^3 = 10^{-6}$, with scattering length $a \approx 5 \times 10^{-7}cm$, trap size $L \approx 5 \times 10^{-4}cm$ and $m = 10^{-25}kg$. The value of the parameter $q$ is restricted, for $q < 1$, by the condition $E_0^{(1)} < E_0^{(0)}$. For the case displayed in Figure 4 the range of allowed values is given by $0.09 < q < \infty$.

# 5 Conclusions

In this paper, we first studied the thermodynamics of the simplest quantum group invariant hamiltonian including a harmonic potential. For $q = 1$, this system becomes a fermion gas with two degrees of freedom in a harmonic potential trap in a $D$-dimensional space. We have analyzed this model by using the representation of the operators $\Psi_i$ in terms of fermion operators.

A calculation of the chemical potential shows that for $q < 1$ it acquires a linear temperature dependent term no present in the fermion, $SU(2)$, case. One consequence of having this linear term is that for $D = 1$ the chemical potential is not constant but decreases with the temperature. At moderate temperatures the heat capacity increases with $q$, and at high temperatures $C = Dk\langle N\rangle$, becoming independently of the value of $q$. A calculation in the Thomas-Fermi approximation at $T = 0$ shows that the particle spatial distribution is independent of $q$ for $q \geq 1$. For $q < 1$, if we decrease the value of $q$, it makes the interaction more repulsive and thus the gas becomes less confined. Since this fermion interaction vanishes for $q = 1$, the depletion of the gas for $q < 1$ is a direct consequence of imposing quantum group invariance. In the last Section we included a fourth-order interaction in terms of the $\Psi$ operators, and calculated the ground state energy up to first order. This fourth-order interaction becomes a sixth-order interaction in terms of fermion operators. A graph of the function $\varepsilon_0$ illustrates that even small deviations from the standard value $q = 1$ have a nontrivial effect on the ground state energy. The ground state energy decreases as the value of $q$ increases and it becomes approximately constant for $q > 5$.

Our main goal has been in studying the phenomenological aspects of a model having the quantum group $SU_q(2)$ as its symmetry group. In the same fashion that a quantum group is considered a deformation of a classical Lie group, we could consider the $SU_q(2)$ symmetry of our model as the result of a small breaking of $SU(2)$ spin symmetry. An alternative view would be to regard $SU_q(2)$ as an additional symmetry independent of the spin degrees of freedom.

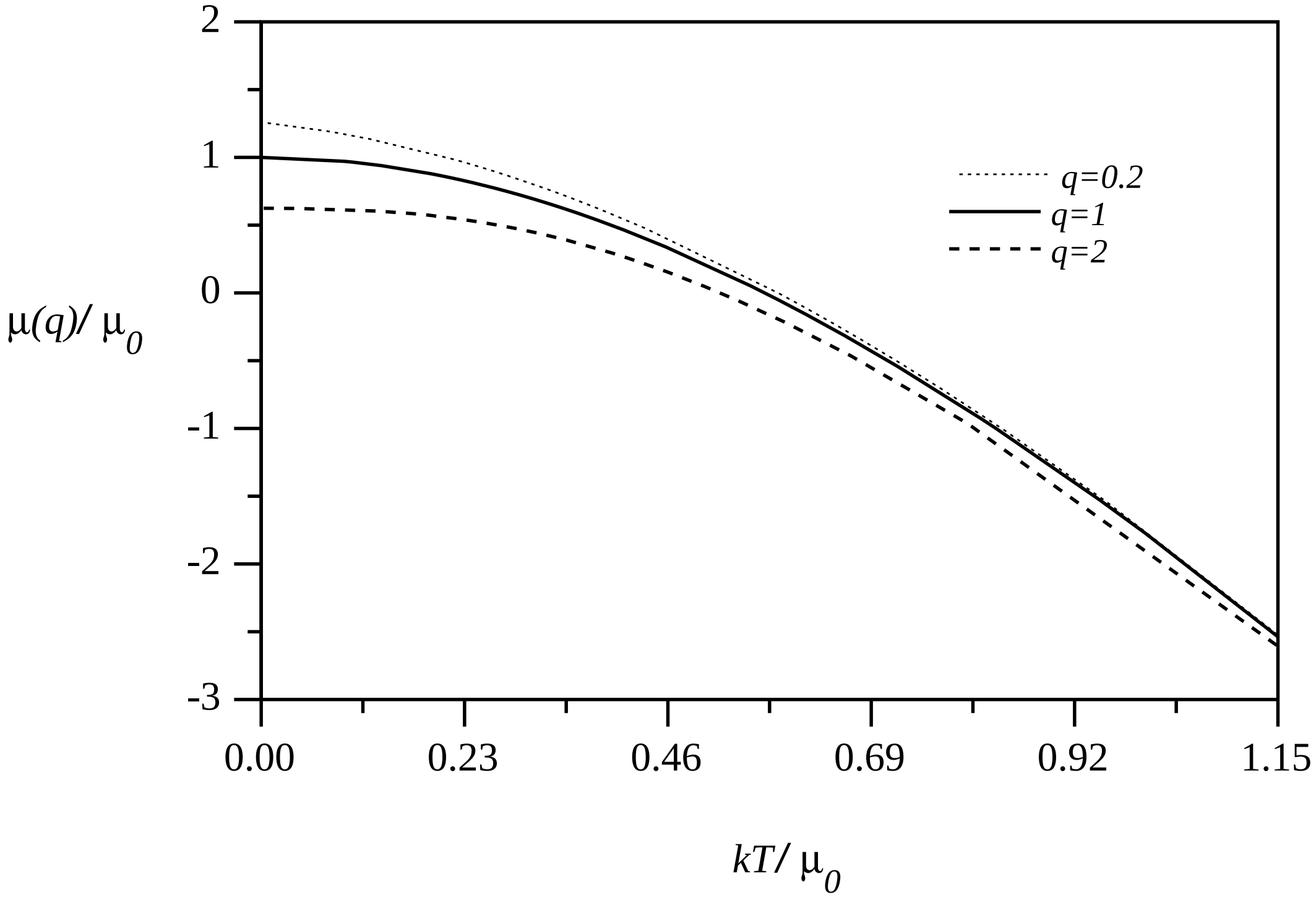

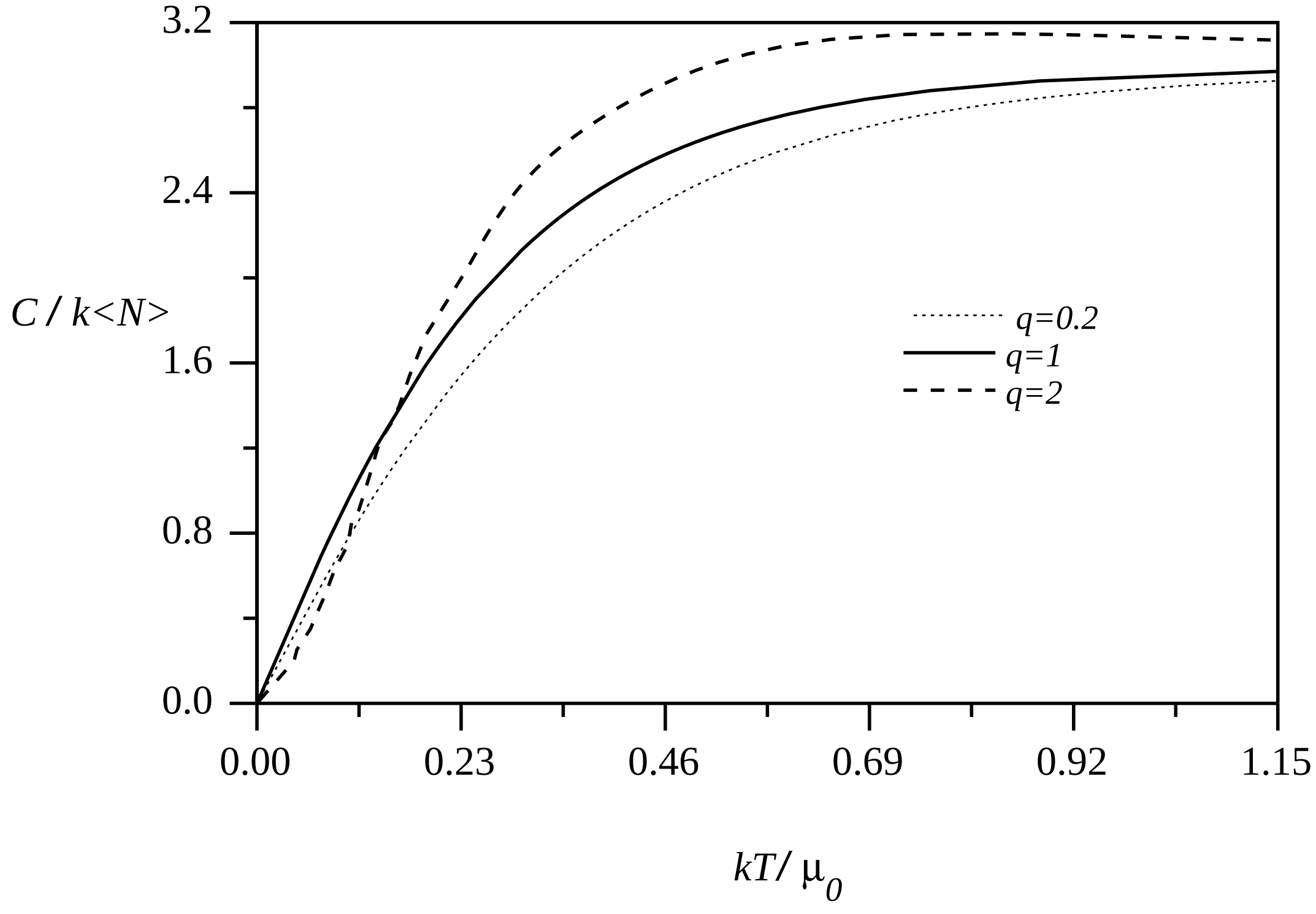

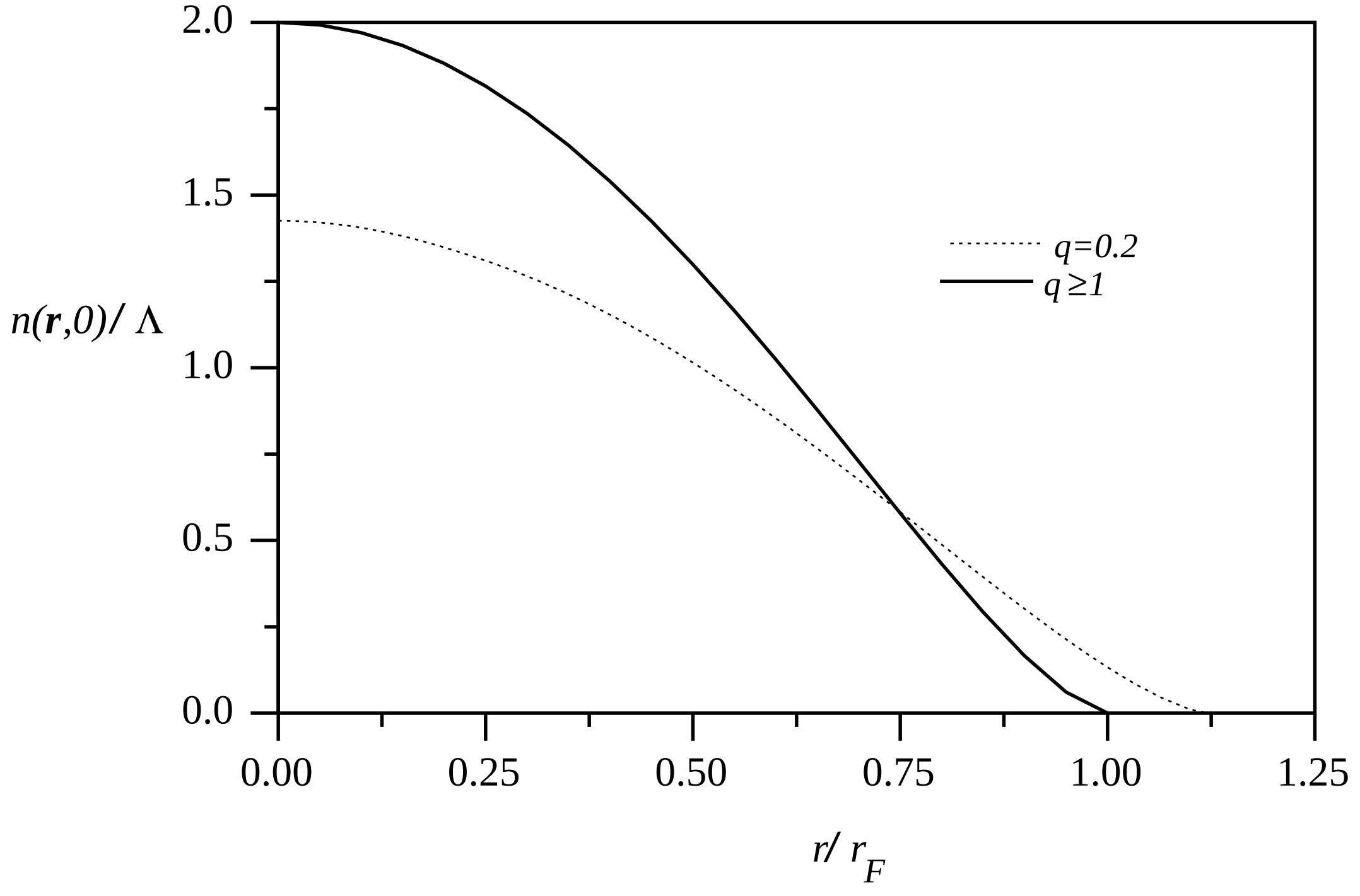

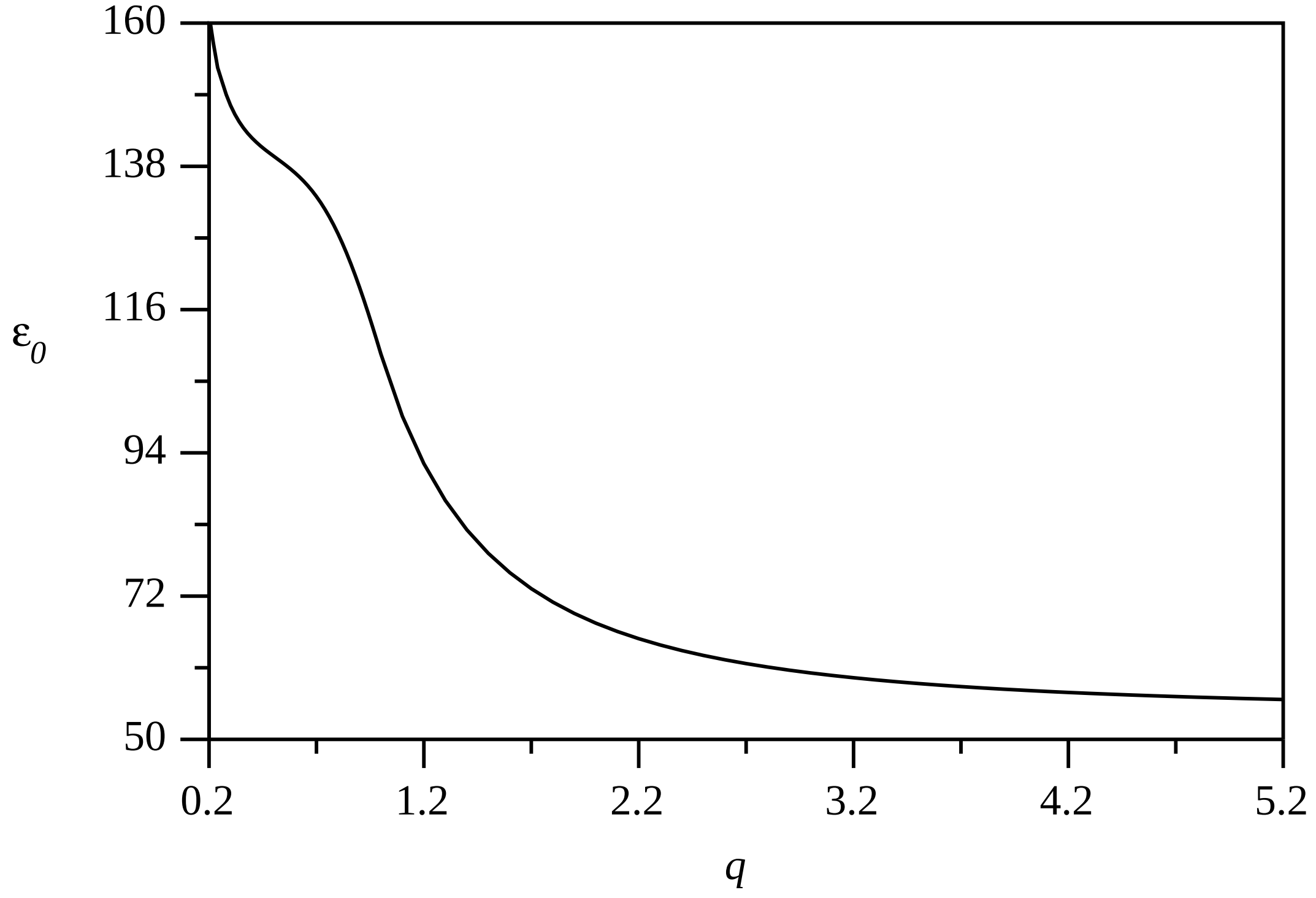

FIG. 1. The chemical potential $\mu(q)$ for $D = 3$ as a function of $T$ for the parameter values $q = 0.2, 1, 2$.

FIG. 2. The heat capacity, as given in Equation (19), for $D = 3$ and the values $q = 0.2, 1, 2$.

FIG. 3. The spatial distribution for $D = 3$ as a function of $r/r_F$.

FIG. 4. The dependence of the function $\varepsilon_0 = 4E_0/g\langle N\rangle^2$ on the parameter $q$ for $na^3 = 10^{-6}$, scattering length $a \approx 5 \times 10^{-7} cm$ and trap size $L \approx 5 \times 10^{-4} cm$.